\documentclass[reprint,aps,prl,floatfix,groupedaddress]{revtex4-2}

\usepackage{graphicx}
\usepackage{amsmath}
\usepackage{mathptmx}
\usepackage{ulem}
\usepackage{xcolor}

\begin{document}

\title{Nonlinear Terahertz Polarizability of Electrons Solvated in a Polar Liquid}
\author{Matthias Runge}
\email{runge@mbi-berlin.de}
\author{Klaus Reimann}
\author{Michael Woerner}
\author{Thomas Elsaesser}
\affiliation{Max-Born-Institut f\"ur Nichtlineare Optik und
Kurzzeitspektroskopie, 12489 Berlin, Germany}

\date{\today}

\begin{abstract}
The nonlinear polaronic response of electrons solvated in liquid 2-propanol is studied by  two-dimensional terahertz spectroscopy. Solvated electrons with a concentration of $c_e \approx 800~\mu$M are generated by femtosecond photoionization of alcohol molecules. Electron relaxation to a localized ground state impulsively excites coherent polaron oscillations with a frequency of 3.9~THz. Off-resonant perturbation of the THz coherence by a pulse centered at 1.5~THz modifies the polaron oscillation phase. This nonlinear change of electron polarizability is reproduced by theoretical calculations.
\end{abstract}

\maketitle

The interaction of an electron with a polar or ionic environment results in the formation of a composite quasi-particle, the polaron \cite{La33}. In crystalline solids, the polaron picture has been applied for describing both the quantum ground state of localized charges and the electron transport in continuum states \cite{FR54A,FH62,PD85}. In the latter, the polaron represents an electron dressed by a phonon cloud, which can undergo a joint center-of-mass motion and/or display internal excitations, e.g., longitudinal elongations of the phonon cloud \cite{FR54A,FH62,PD85}. The linear and nonlinear response of polarons to external electric fields has mainly been elucidated in studies of charge transport in solids by femtosecond all-optical experiments \cite{Ba93,JZ95,SW03,BG01,GK07}.

Much less is known on polaron excitations in systems without a periodic long-range order such as polar liquids \cite{CS84,MP87}. In this context, solvated electrons represent a prototypical quantum system. In equilibrium, they populate the ground state in a self-consistent potential well, which has been described by polaron models in analogy to a localized charge in a crystal \cite{SC60, LA91, WE72}, as a quantum state at a site of enhanced solvent density \cite{LA10}, or as the ground state in a local void or cavity in the liquid \cite{TU12B}. The widely accepted cavity model accounts for a broad range of properties of electrons solvated in hydrogen-bonding liquids such as water and alcohols  \cite{TU12B,WE91,KE82,SH95,MI99,WA16}.  In water and methanol, OH groups of the four to six solvent molecules in the first solvation shell point towards the electron, thus minimizing the electrostatic energy and confining the electron wavefunction to a  radius of gyration of 2.5~\AA~(water) and  2.2~\AA~(methanol).

The long-range Coulomb interaction of the confined electron with the solvent results in a coupling of electronic excitations and low-frequency molecular motions in the liquid, in analogy to electron-phonon coupling in a crystal. 
First evidence for such polaronic excitations in water and alcohols originates from very recent studies of their ultrafast THz response. In a novel approach,  solvated electrons have been generated by tunneling ionization in the fluctuating electric field of liquid water and separation of electron and parent ion in a strong THz field \cite{GH20A}. The presence of solvated electrons results in pronounced changes of the real and imaginary part of the THz dielectric function of water and alcohols via the electron polarizability \cite{GH20A,SI22B,WO22}. Upon femtosecond electron relaxation from delocalized continuum states into the localized ground state, collective polaron oscillations are excited impulsively, giving evidence of a transient many-body response on a length scale set by the Debye screening length of the electron's electric field \cite{GH21B,SI22B}.

Polaron oscillations are connected with a radial charge density modulation and, thus, are purely longitudinal. Depending on the charge density distribution within the polaron, the Debye screening length is modified. The resulting periodic modification in polaron size creates a transverse polarizability through which the polaron can be accessed with transverse optical fields \cite{WO22}. 
While time-resolved THz spectroscopy has so far focused on the linear polaron response, there are first indications of a nonlinear interaction of THz electric fields with polarons in alcohols \cite{SI22B}. However, a systematic experimental or theoretical study of  polaronic nonlinearities does not exist. Here, multidimensional THz spectroscopy holds a particular potential for revealing  nonlinearities in the electric polarizability of the liquid.

In this Letter, we present new insight in the nonlinear THz response of polarons in 2-propanol (isopropanol, IPA). Fully phase-resolved two-dimensional THz (2D-THz) experiments on solvated electrons generated by multiphoton excitation reveal a strong nonlinear change of THz polarizability. Coherent polaron oscillations, which are launched upon electron localization, display pronounced changes of oscillation phase under the action of a non-resonant THz pulse. This nonlinear THz response is reproduced by calculations introducing a nonlinear THz polarizability of the solvated electrons. 

\begin{figure}[t]
	\begin{center}
		\includegraphics[width=0.5\textwidth]{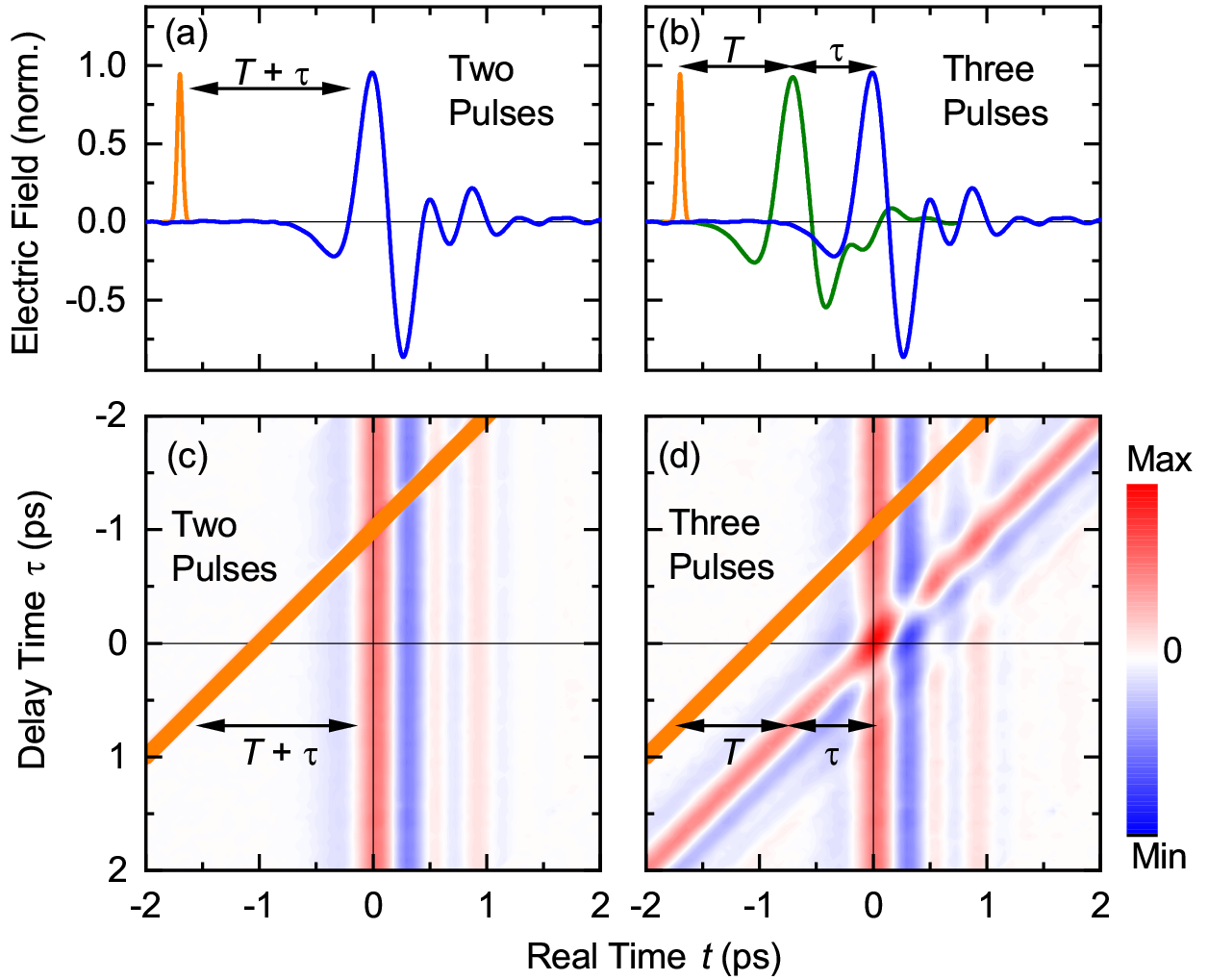}
	\end{center}
	\caption{Experimental concept and 2D-THz data at a waiting time of $T=1~$ps. 
			(a), (b) Two- and three-pulse sequences consisting of a femtosecond NIR pulse, and one or two (phase-locked) THz pulses as a function of real time $t$. The time intervals $T$ and $\tau$ are the waiting and delay time. 
			(c) Contour plot of $E^{c_e}_{\text{Pr}}(t,\tau)$ (two-pulse experiment) with the NIR pump pulse (orange line) and THz probe pulse $E^{0}_{\text{Pr}}(t)$ (vertical trace). (d) Contour plot of 
			$E^{c_e}_{\text{both}}(t,\tau)$ (three-pulse experiment) with the additional perturbing THz pulse $E^{c_e}_{\text{Pert}}(t,\tau)$ (diagonal trace). 
	}   
	\label{fig:setup}
\end{figure}

The experiments are based on two- and three-pulse sequences  [Figs.~\ref{fig:setup}(a) and (b)], consisting of a femtosecond near-infrared (NIR) pulse for electron generation and one or two phase-locked THz pulses for mapping the nonlinear response of solvated electrons. Solvated electrons are generated in a liquid jet of IPA by multiphoton ionization with an 800-nm pulse, providing electron concentrations of up to $c_e\approx800~\mu$M. Relaxation of the photo-generated electrons to their localized ground state initiates, in an impulsive way, coherent underdamped polaron oscillations in the liquid \cite{GH21B,SI22B}.  Details of the experimental setup, choice of the solvent, and  electron generation are given in the Supplementary Material (SM) \cite{SM}.

In the experiments with the two-pulse sequence, the THz pulse maps the polaron response of the sample arising after electron generation by the NIR pump pulse, in particular the impulsively excited coherent polaron oscillations. Such data serve as a benchmark for the results of the three-pulse experiments, in which the polaron oscillations are perturbed by interaction with the first THz pulse after  waiting times of $T=~1$ or 75~ps. The impact of this interaction is probed by the second THz pulse at a time delay $\tau$ relative to the first THz pulse. The two THz pulses have peak electric fields of 50 kV/cm and leave the electron concentration $c_e$ unchanged  \cite{GH20A}. 
The transmitted THz pulses are detected by electrooptic sampling, i.e., their electric field is measured as a function of real time $t$.

The nonlinear signal generated in the two-pulse sequence is given by $E_{\text{NL}}(t,\tau)=E^{c_e}_{\text{Pr}}(t,\tau)-E^{0}_{\text{Pr}}(t)$. Here $E^{c_e}_{\text{Pr}}(t,\tau)$ is the THz probe field transmitted after electron generation, and $E^{0}_{\text{Pr}}(t)$ is the transmitted field of the THz probe pulse without solvated electrons. For the three-pulse sequence, the nonlinear signal is $E^{\text{Pert}}_{\text{NL}}(t,\tau)=E^{c_e}_{\text{both}}(t,\tau)-E^{c_e}_{\text{Pert}}(t,\tau)-E^{0}_{\text{Pr}}(t)$, where $E^{c_e}_{\text{both}}(t,\tau)$ represents the electric field of the two THz pulses transmitted through the sample after electron generation, and $E^{c_e}_{\text{Pert}}(t,\tau)$ the transmitted field of the perturbing first THz pulse after electron generation. In all measurements, a weak THz signal directly generated by the 800-nm pulse with an amplitude of 0.3 kV/cm was subtracted from the transmitted THz fields. More details of the experimental setup and 2D-THz spectroscopy are given in the SM and in Ref.~\cite{RE21A}.

\begin{figure}[t]
	\begin{center}
		\includegraphics[width=0.5\textwidth]{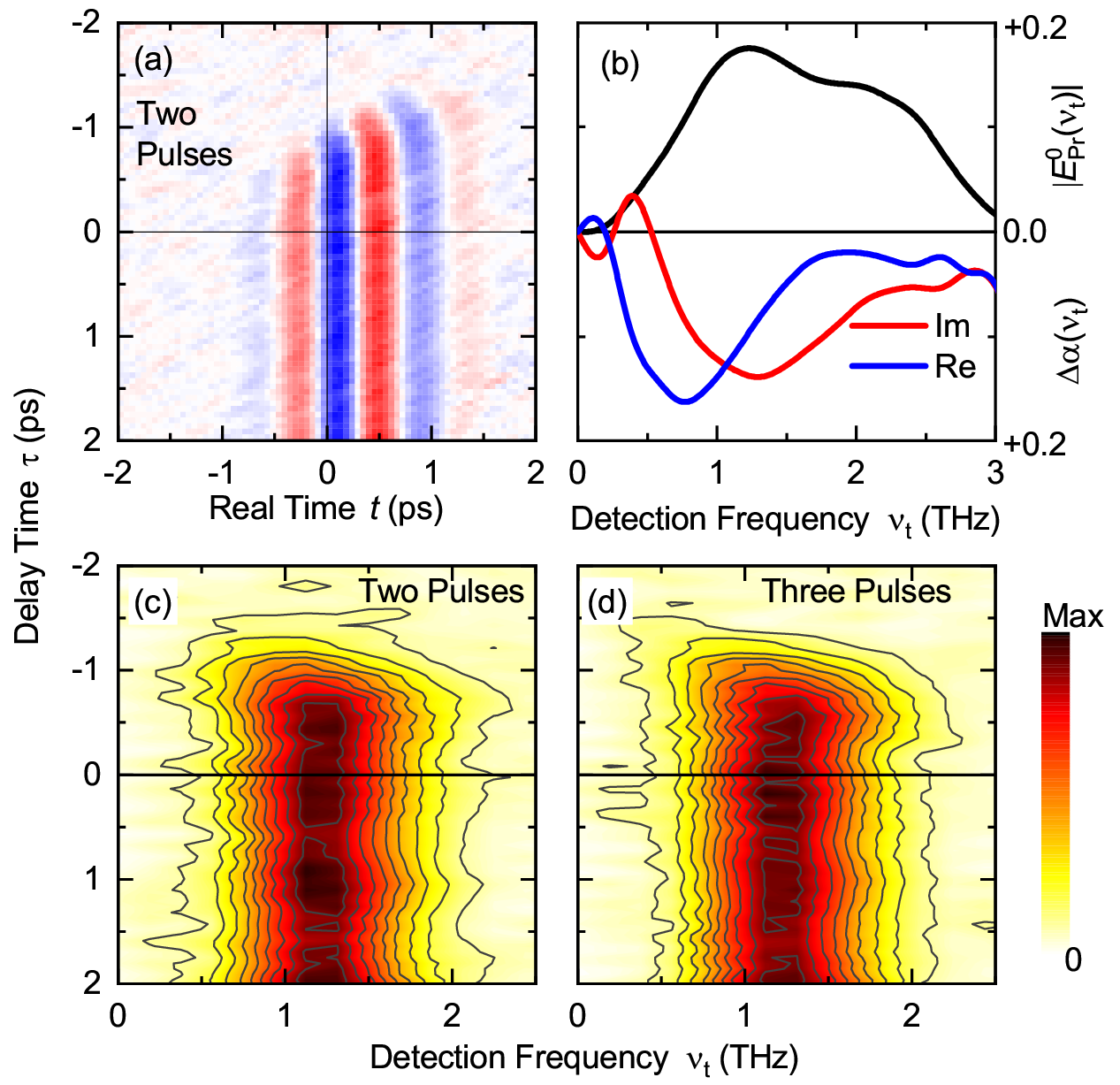}
	\end{center}
	\caption{Nonlinear 2D-THz data at a waiting time $T=1~$ps. 
		(a)~Contour plot of the nonlinear 2D-THz signal $E_{\text{NL}}(t,\tau)$.
	 (b)~Probe pulse spectrum (black line) and real and imaginary parts (blue and red lines) of the spectrally resolved pump-probe signal $\Delta \alpha(\nu_t,\tau_{\text{avg}})$ averaged over delay times from $\tau_{\text{avg}}=-0.75$~to~$2$~ps. 
		(c) Contour plot of $|E_{\text{NL}}(\nu_t,\tau)|$ obtained from a Fourier transform of  $E_{\text{NL}}(t,\tau)$.
	(d) Contour plot of $|E^{\text{Pert}}_{\text{NL}}(\nu_t,\tau)|$, the Fourier transform of  $E^{\text{Pert}}_{\text{NL}}(t,\tau)$ (not shown).	
	}
	\label{fig:2Dscan}
\end{figure}

In the following, we present two sets of data recorded for waiting times $T=1$~ps and $T=75$~ps.  In Figs.~\ref{fig:setup}(c) and (d), the electric fields $E^{c_e}_{\text{Pr}}(t,\tau)$ (single THz pulse) and $E^{c_e}_{\text{both}}(t,\tau)$ (two THz pulses) are plotted as a function of real time (abscissa) and delay time (ordinate). From the two-pulse data in panel (c),  the unperturbed THz response $E_\text{NL}(t,\tau)$ of the solvated electrons for $T=1$~ps is extracted [Fig.~\ref{fig:2Dscan}(a)]. The signal exhibits an absorption increase with an amplitude of up to 10~kV/cm, reflecting the change of the macroscopic dielectric function induced by electron generation.

Figure~\ref{fig:2Dscan}(b) shows the probe spectrum $|E^0_{\text{Pr}}(\nu_t)|$ together with the real and imaginary parts of the two-pulse pump-probe signal $\Delta \alpha (\nu_t,\tau_{\text{ave}}) = - \text{ln}[E^{c_e}_{\text{Pr}}(\nu_t,\tau_{\text{ave}})/E^0_{\text{Pr}}(\nu_t,\tau_{\text{ave}})]$, averaged over delay times from $\tau_\text{avg}=~-0.75$~to~2~ps. One observes an absorption increase with a maximum at $\nu_t=0.8$~THz and pronounced changes of the THz refractive index, indicating the spectral region most sensitive to changes in the macroscopic polarizability. 
In Fig.~\ref{fig:2Dscan}(c), the spectrum $|E_{\text{NL}}(\nu_t,\tau)|$ derived by Fourier transforming $E_{\text{NL}}(t,\tau)$  along  $t$ is plotted as a function of detection frequency $\nu_t$ and $\tau$. For noise reduction, the data were Fourier-filtered by a Gaussian filter of 10~THz width in excitation frequency $\nu_\tau$ and of 2~THz width in detection frequency $\nu_t$ (FWHM) at a spectral position of $(\nu_t,\nu_\tau)=(1.5,0)$~THz.
The low-frequency spectral wing of $|E_{\text{NL}}(\nu_t,\tau)|$  shows a distinct oscillatory behavior, as is evident from the shape of the contour lines. 

\begin{figure}[t]
	\begin{center}
		\includegraphics[width=0.5\textwidth]{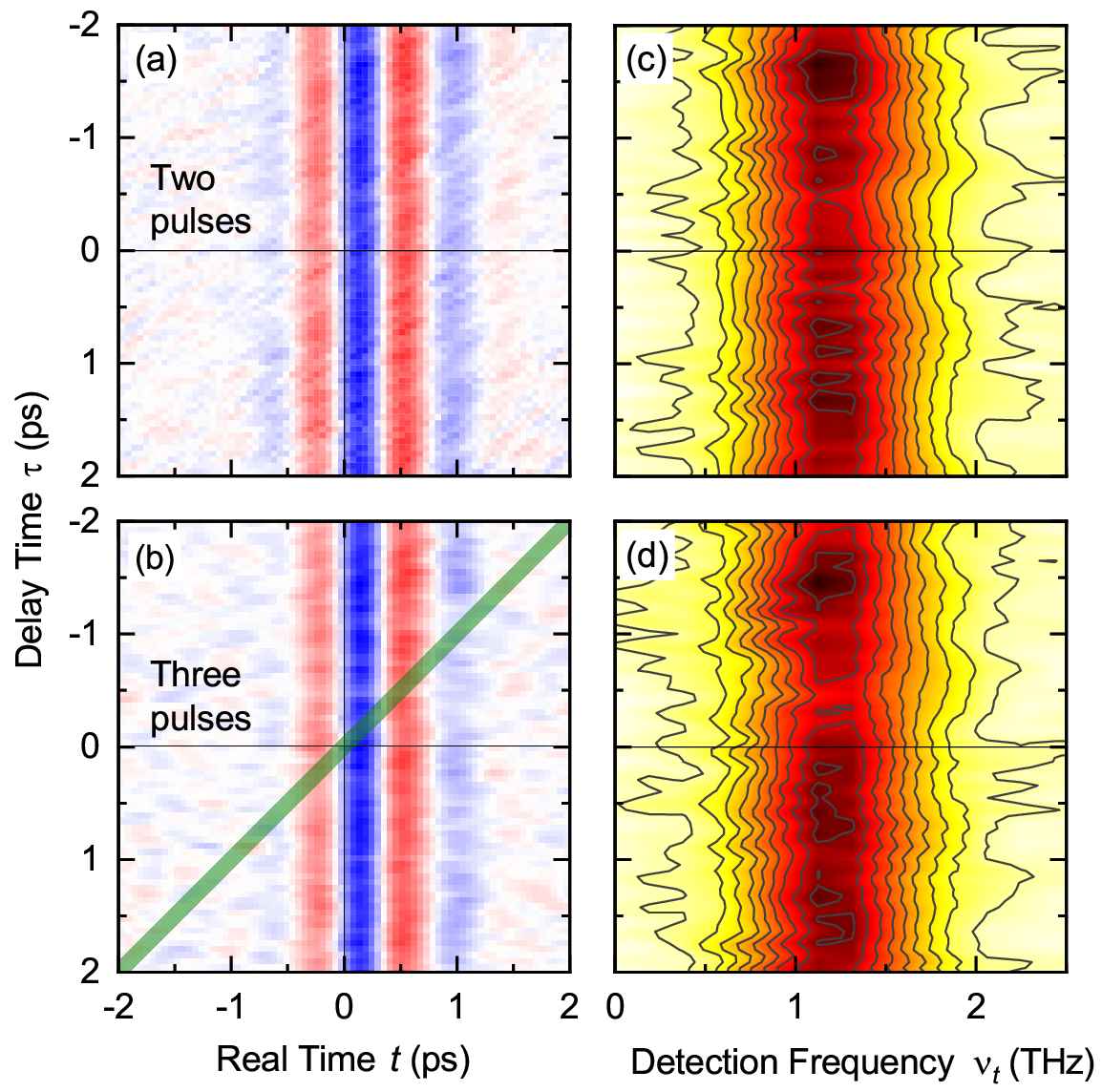}
	\end{center}
	\caption{2D-THz data recorded at $T=75$~ps. (a),~(b)~Contour plots of the 2D-THz signals $E_{\text{NL}}(t,\tau)$ and $E^{\text{Pert}}_{\text{NL}}(t,\tau)$ for the two- and three-pulse configuration. The green line in panel (b) indicates the temporal position of the perturbing THz pulse.
		(c),~(d)~Contour plots of $|E_{\text{NL}}(\nu_t,\tau)|$ and $|E_{\text{NL}}^{\text{Pert}}(\nu_t,\tau)|$  as functions of detection frequency $\nu_t$ and delay time $\tau$.
	}
	\label{fig:2Dscan_75}
\end{figure}

Results of the three-pulse experiment with a waiting time $T=1$~ps [Fig.~\ref{fig:setup}(d)] are presented in Fig.~\ref{fig:2Dscan}(d), showing the spectrum $|E^{\text{Pert}}_{\text{NL}}(\nu_t,\tau)|$. The line shape is similar to the two-pulse spectrum in  Fig.~\ref{fig:2Dscan}(c). However, there are changes in the oscillatory response around $\tau  = 0$ that will be analyzed below. In Figs.~\ref{fig:2Dscan_75}(a)~and~(b), $E_{\text{NL}}(t,\tau)$ and $E_{\text{NL}}^\text{Pert}(t,\tau)$ are shown for $T=75~$ps, while  Figs.~\ref{fig:2Dscan_75}(c)~and~(d) display the corresponding spectra  $|E_{\text{NL}}(\nu_t,\tau)|$ and $|E^{\text{Pert}}_{\text{NL}}(\nu_t,\tau)|$. The qualitative behavior is similar to the data for $T=1$~ps (Fig.~\ref{fig:2Dscan}), a result of the underdamped character of the polaron oscillations, which persist for delay times $\tau$ well beyond 100~ps.

\begin{figure}[t]
	\begin{center}
		\includegraphics[width=0.5\textwidth]{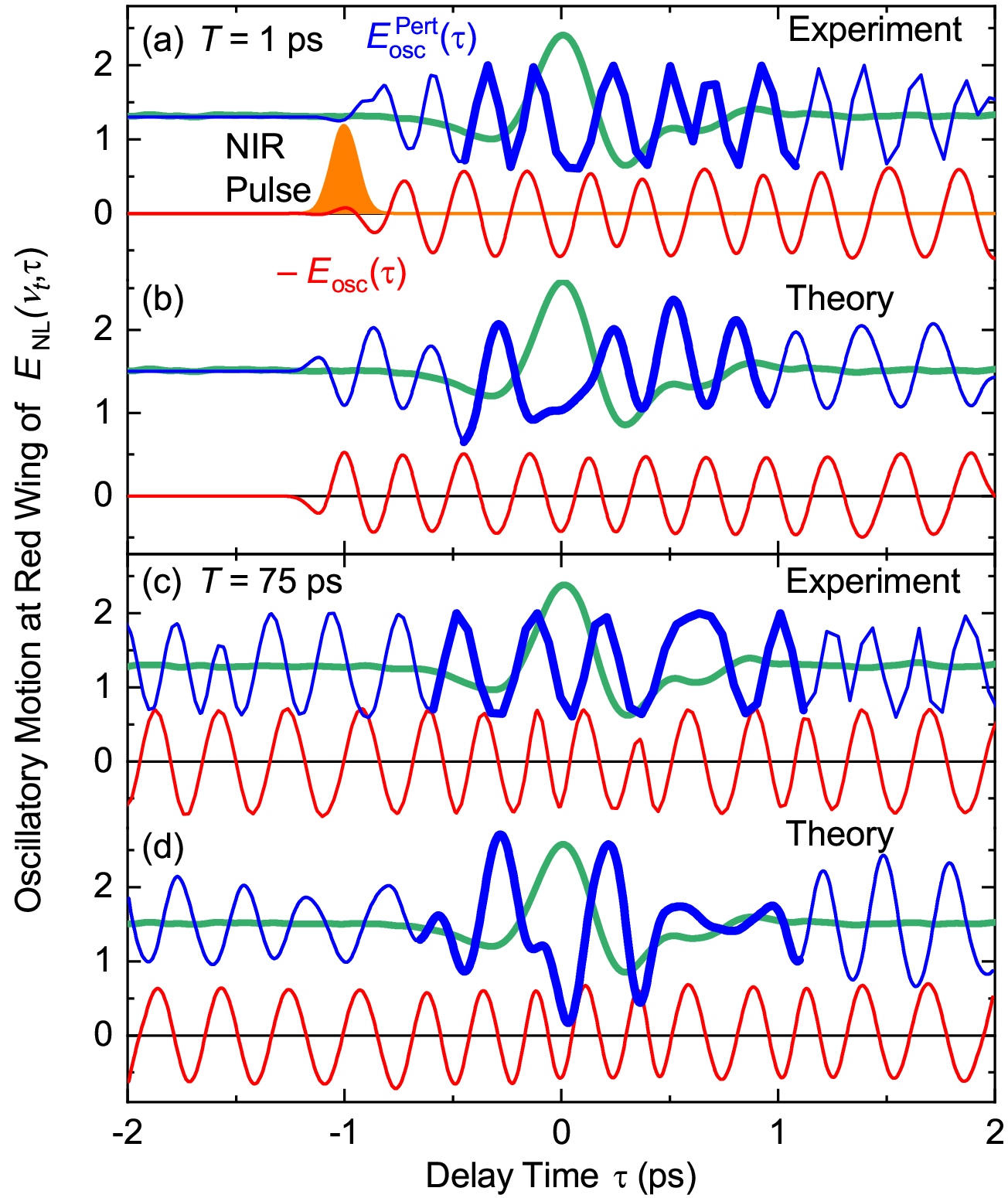}
	\end{center}
	\caption{(a),~(c)~Oscillatory signal components $E^\text{Pert}_{\text{osc}}(\tau)$ (blue lines) and $-E_{\text{osc}}(\tau)$ (red lines) with and without external THz perturbation for $T=1$ ps and $T=75$ ps. The thick blue lines mark the time range of THz perturbation. (b),~(d)~Transients calculated from the theoretical model for $T=1$ ps and $T=75$ ps. In all panels, the green lines give the perturbing THz field.}
	\label{fig:plasosc}
\end{figure}   

To characterize the oscillatory nonlinear response and to identify the impact of the perturbing THz pulse on the polaron oscillations, we
analyze the low-frequency wings of the spectra for $T=1$~ps [Figs.~\ref{fig:2Dscan}(c) and (d)] and $T= 75$~ps   [Figs.~\ref{fig:2Dscan_75}(c) and (d)] with a method detailed in the SM. In brief, cuts of the pump-probe signal along delay time $\tau$ in a spectral window  from $\nu_t=0.5$ to $0.9$~THz are fitted by a sine function with a $\tau$-dependent amplitude and phase, plus a background with a $\tau$-dependent amplitude. The oscillatory signals derived by subtracting the background are plotted in Fig.~\ref{fig:plasosc}(a) and (c) for $T=1~$ps and 75~ps, respectively. Here, red lines represent the field $-E_{\rm osc}(\tau)$ from the two-pulse experiments, while blue lines give the field $E_{\rm osc}^{\rm Pert}(\tau)$ from the three-pulse experiments. As a reference, the electric field of the perturbing THz pulse is shown (green line). The oscillation frequency of $-E_{\rm osc}(\tau)$ traces is 3.9 THz, which represents the polaron frequency for $c_e \approx 800$ $\mu$M.  
For $\tau<-0.5$~ps in panel (a), $-E_\text{osc}(\tau)$ and $E_\text{osc}^\text{Pert}(\tau)$ display a constant relative phase shift of $\pi$, i.e., $+E_\text{osc}(\tau)$ and $E_\text{osc}^\text{Pert}(\tau)$ are in phase. From $\tau=-0.5$~to~1~ps in panel (a), $E_\text{osc}(\tau)$ does not change, whereas the phase of $E_\text{osc}^\text{Pert}(\tau)$ is strongly modified. This phase change reflects a momentary change of polaron frequency induced by the perturbing nonresonant THz pulse, a novel type of nonlinear response. Panel (c) shows a very similar behavior at $T=75$~ps.  In the SM, we provide a more detailed discussion of the properties of the momentary perturbed phase.

The oscillatory pump-probe signals are due to coherent polaron oscillations, induced impulsively during the subpicosecond electron localization process \cite{GH21B,SI22B}.  Each electron couples to electronic and nuclear degrees of freedom of the polar environment, thus forming a collective polaron excitation. In space, the relevant  coupling range is roughly set by the Debye screening length of $L_D=1.8$~nm, which includes some 190 IPA molecules and is much smaller than the average distance between electrons of 13~nm. These values were estimated from the real part of the dielectric function $\text{Re}[\epsilon_\text{IPA}(1~$THz$)]=2.25$ \cite{SM} and  $c_e=800~\mu$M. 
The polaron oscillations are longitudinal coherent motions of space charge within a sphere of radius $L_D$, which primarily arise from the superposition of orientational motions of IPA dipoles.  The longitudinal electric field is fully screened by a surface charge on the shell of the sphere \cite{FR54A}. The polaron (oscillation) frequency is determined by the zero crossing of the real part of the longitudinal dielectric function inside the sphere \cite{FU68A}. The longitudinal oscillations of space charge modulate the radius of the sphere, in this way affecting the macroscopic \textit{ transversal} polarizability of the liquid. In the SM, we compare experimental polaron frequencies for a wide range of electron concentration with calculations based on a Clausius-Mossotti local field picture  \cite{HA83A,GH20A}.

The nonresonant perturbing THz pulse induces a nonlinear polarization acting on the polaron. This nonlinear response causes the observed phase  changes  in the polaron oscillations (cf. Fig.~\ref{fig:plasosc}). The modulations prevail in the time window of interaction with the perturbing pulse. Even after the interaction, however, the initially fixed phase relation between polaron oscillations with and without perturbing THz field is somewhat softened. This effect may arise from a correlation of polarizations in the system. The transverse polarization correlation function determines the dephasing of transversal excitations and accounts for correlations in electric currents. The phase deviations after the perturbing THz pulse suggest correlations persisting on a time scale longer than the 1-ps duration of the perturbing pulse \cite{EL16A}.

To account for the nonlinear polaron response, we present a theoretical model based on a Clausius-Mossotti approach \cite{HA83A}. The total electric polarization 
\begin{equation}
P_\text{tot}= P^{\text{el}}_{\text{bound}} + P^{\text{nuc}}_{\text{Debye}} +  P^{\text{el}}_{\text{free}}\label{eq:Ptot}
\end{equation}
includes contributions from all spatially disjunct dipoles, i.e., $P^{\text el}_{\text{bound}}$ from dipoles generated by bound electrons, $P^{\text{nuc}}_{\text{Debye}}$ for molecular dipoles related to nuclear motions, and $P^{\text{el}}_{\text{free}}$ for electronic dipoles caused by free electron motions. Such dipoles are mutually coupled  via the local electric field, consisting of the macroscopic electric field $E(t)$ and $P_\text{tot}/(3\epsilon_0)$. This approach leads to the equations  
\begin{eqnarray}
	3\frac{\epsilon_\text{IPA}(\omega)-1}{\epsilon_\text{IPA}(\omega)+2}&=& \chi_{\text{hf}} +\frac{\chi_{\text{lf}}-\chi_{\text{hf}}}{1+i\omega\tau_R}\label{eq:CM}\\
	P^{\text{el}}_{\text{bound}}&=&\epsilon_0\chi_{\text{hf}}\left(E(t)+\frac{P_{\text{tot}}}{3\epsilon_0}\right)\label{eq:hf}\\
	\frac{d P^{\text{nuc}}_{\text{Debye}}}{dt}&=&-\frac{1}{\tau_R}P^{\text{nuc}}_{\text{Debye}}+\epsilon_0\frac{\chi_{\text{lf}}-\chi_{\text{hf}}}{\tau_R}\left(E(t)+\frac{P_{\text{tot}}}{3\epsilon_0}\right)\label{eq:Debye}\\
	\frac{d^2 P^{\text{el}}_{\text{free}}}{dt^2}&=& -\gamma \frac{d P^{\text{el}}_{\text{free}}}{dt} - \left(N_{\text{el}} +N_{\text{osc}}\right)\frac{e_0^2}{m_e} E_\text{cr} \times \nonumber\\
	& &\sinh\left(\frac{1}{ E_{\text{cr}}}\left[E(t)+\frac{P_{\text{tot}}}{3\epsilon_0 }\right] \right)\hspace{5mm}\label{eq:free}\\
	\frac{d N_\text{el}}{dt}&=&A\cdot I_\text{NIR}\label{eq:Nel}\\ 
	\frac{d^2 N_\text{osc}}{dt^2}&=& - B(N_\text{el})\cdot N_\text{osc} + A\cdot C \cdot I_\text{NIR}\label{eq:Nosc}\\
	E_\text{em}&=&-\frac{d}{2\epsilon_0 c}\cdot \frac{d P_\text{tot}}{dt}\label{eq:Eem}
\end{eqnarray}	
The high and low frequency susceptibilities $\chi_{\text{hf}}$ and $\chi_{\text{lf}}$, and the Debye relaxation time $\tau_\text{R}$ of neat IPA are derived from a fit of $\epsilon_\text{IPA}(\omega)$ to linear THz spectra \cite{SM}. The electric field is given by 
$E(t)=E_{\text{pert}}(t,\tau)+E^0_{\text{pr}}(t)$ of the two experimental THz pulses [cf.~Fig.~\ref{fig:setup}(b)].  
The electron density $N_{\text{el}}=c_eN_\text{A}$ ($N_A$: Avogadro constant) and the amplitude $N_{\text{osc}}$ of longitudinal electron density oscillations are given in \eqref{eq:Nel} and \eqref{eq:Nosc}, where $I_{\text{NIR}}$ is the intensity of the electron generating NIR pulse. We treat the influence of longitudinal polaron oscillations \eqref{eq:Nosc} on the macroscopic transverse polarization as density oscillations  \eqref{eq:free}, rather than considering an oscillating polaron size  modulating the polarizability. $\gamma$ in (5) is the damping constant of the transverse polarization. The parameters $A$, $C$, and the function $B(N_\text{el})$ are chosen to reproduce the experimental electron density, polaron oscillation amplitude and polaron frequency. The emitted electric field \eqref{eq:Eem} is determined by $P_\text{tot}$ and the sample thickness $d$ ($\epsilon_0$: vacuum permittivity; $c$: speed of light). Without generation of solvated electrons ($N_\text{el}=N_\text{osc}=0$), the THz response defined by \eqref{eq:hf} and \eqref{eq:Debye} is strictly linear. The only nonlinearity considered here is the nonlinear polarizability of the solvated electrons described by the sinh term in \eqref{eq:free} with a strength characterized by a critical electric field $E_\text{cr}$. Slight drifts of the experimental polaron frequency (cf. red time traces in Fig.~\ref{fig:plasosc}), which originate from fluctuations in the density of photogenerated electrons, are taken into account in the calculations \cite{SM}.

Results of the calculations are presented in Figs.~\ref{fig:plasosc}(b) and~(d) for $T=1~$ps and $T=$ 75~ps. In analogy to the experimental result [Figs.~\ref{fig:plasosc}(a) and (c)], oscillatory electric field changes on the red wing of the THz emission spectrum $E_{\text{em}}(\nu_t, \tau)$ are plotted as a function of delay time $\tau$. The experimental characteristics are best reproduced with $E_\text{cr}=100$~kV/cm, for which the oscillations show distinct phase and amplitude modulations in the time range of the external perturbation. In addition to changes of the momentary phase, the calculations suggest increased amplitudes during interaction with the perturbing THz pulse. With our method applied for data analysis \cite{SM}, we have no direct access to amplitude variations.

In conclusion, solvated electrons in a polar liquid show a polaronic nonlinear response induced by a nonresonant THz pulse. This novel effect is mapped via distinct phase modulations of coherent underdamped polaron oscillations at a high polaron frequency of 3.9~THz. Our findings underline the relevance of many-body excitations in polar molecular ensembles and represent a pathway of modifying their dielectric properties by external THz fields.

\begin{acknowledgments}
This research has received funding from the European Research Council (ERC) under the European Union’s Horizon 2020 Research and Innovation Program (grant
agreement 833365).
\end{acknowledgments}

\bibliographystyle{nature2}

\ifx\unpublished\@undefined\def\unpublished{unpublished}\fi

\end{document}